\documentclass[preprint]{aastex}
\usepackage{url}
\usepackage{hyperref}

\shorttitle{Tornado Prominence}
\shortauthors{Wang et al.}

\begin{document}

\title{Tornado-Like Evolution of A Kink-Unstable Solar Prominence}

\author{Wensi Wang\altaffilmark{1}, Rui Liu\altaffilmark{1,2}, Yuming Wang\altaffilmark{1,3}   }

\altaffiltext{1}{CAS Key Laboratory of Geospace Environment, Department of Geophysics and Planetary Sciences, University of Science and Technology of China, Hefei 230026, China; rliu@ustc.edu.cn}

\altaffiltext{2}{Collaborative Innovation Center of Astronautical Science and Technology, Hefei 230026, China}

\altaffiltext{3}{Synergetic Innovation Center of Quantum Information \& Quantum Physics, University of Science and Technology of China, Hefei 230026, China}

\begin{abstract}
We report on the tornado-like evolution of a quiescent prominence on 2014 November 1. The eastern section of the prominence first rose slowly transforming into an arch-shaped structure as high as $\sim\,$150 Mm above the limb; the arch then writhed moderately in a left-handed sense, while the originally dark prominence material became in emission in the \ion{Fe}{9} 171~{\AA} passband, and a braided structure appeared at the eastern edge of the warped arch. The unraveling of the braided structure was associated with a transient brightening in EUV and apparently contributed to the formation of a curtain-like structure (CLS). The CLS consisted of myriads of thread-like loops rotating counterclockwise about the vertical if viewed from above. Heated prominence material was observed to slide along these loops and land outside the filament channel. The tornado was eventually disintegrated and the remaining material flew along a left-handed helical path of approximately a full turn, as corroborated through stereoscopic reconstruction, into the cavity of the stable, western section of the prominence. We suggest that the tornado-like evolution of the prominence was governed by the helical kink instability, and that the CLS formed through magnetic reconnections between the prominence field and the overlying coronal field.

\end{abstract}


\section{INTRODUCTION}

Solar prominences are cool and dense `clouds' suspended in the hot and tenuous corona. In H$\alpha$, prominences appear in emission above the limb, but in absorption against the disk (also termed \emph{filaments} in this case, but often used interchangeably with \emph{prominences}); in EUV, they generally have a dark appearance, but may become brightened from time to time due to heating, e.g., the interface region between the prominence and the ambient corona often emits in UV/EUV \citep{Labrosse2010}. A `dipped' or helically coiled field can support prominence material against gravity \citep[e.g.,][]{ks1957,kr1974,Low&Hundausen1995}.  Alternatively, a steady-state dynamic solution in flat-topped arcade fields is able to explain the formation and suspension of prominences through thermal non-equilibrium \citep[][and references therein]{Karpen2015}. In particular, the helical flux-rope model makes a favorable comparison with the prominence-cavity system (PCS), in which the prominence is surrounded by the cavity, an elliptical region of closed loops in limb observations \citep{Gibson2015}. 

Being eruptive or not, prominences are dynamic on a wide range of spatio-temporal scales, which is probably associated with plasma instabilities \citep[e.g.,][]{Ryutova2010}. Among them, the helical kink instability has received a lot of attention, mainly motivated by the warped axes of eruptive filaments \citep[][and references therein]{Liu2016}. The instability is triggered when the magnetic twist of a flux rope \citep[winding of magnetic field lines around the rope axis; see also][]{Liu2016} reaches a threshold, typically exceeding one full turn, and is abruptly converted to magnetic writhe (winding of the rope axis itself). Recently, \citet{Hassanin&Kliem2016} conducted a parametric MHD simulation study modeling confined eruptions driven by the helical kink instability. They identified two distinct magnetic reconnection processes in the evolution of a kink-unstable flux rope, i.e., the dissolution of the erupting flux via reconnection with overlying flux, followed by reconnection in the vertical current sheet between the two legs of the original flux rope; consequently a far less twisted flux rope is reformed. 

Non-eruptive prominences that exhibit long-lasting cyclonic behaviors are also termed ``solar tornadoes'' or ``tornado prominences''. In the early studies, they were described as ``vertical spirals or tightly twisted ropes'' \citep[][p15]{Pettit1932}. Recently, \citet{Li2012} reported a tornado prominence characterized by long-lasting swirling motions in a PCS above the limb on 2011 September 25, which was interpreted as mass flows along the helical field of the cavity. Alternatively, \citet{Panesar2013} suggested that this tornado  could be a dynamical response of the helical prominence field to the cavity expansion due to magnetic ``implosion''  \citep{Hudson2000,Liu2009} in the neighboring active region; \citet{LiuJ2012} interpreted the same event as slow magnetoacoustic wave trains traveling in a writhed flux tube; \citet{Panasenco2014} argued that plasma motions along writhed field lines may appear to swirl in projection. Nevertheless, the mass flows are believed to trace the magnetic field of the PCS because of the low-$\beta$ environment in which it resides.

Here we report a solar tornado that exhibits some `fingerprints' of the helical kink instability. Unlike the 2011 September 25 event, most of rotational motions in the present case were about the vertical, which is reminiscent of tornadoes on Earth and in accordance with the writhing motion of the prominence body as a whole. The observation is described and analyzed in detail in Section 2. Concluding remarks are given in Section 3.

\section{OBSERVATIONS AND ANALYSIS}

\subsection{Instruments}
The tornado prominence was observed by the Atmospheric Imaging Assembly \citep[AIA;][]{Lemen2012} onboard the Solar Dynamics Observatory \cite[SDO;][]{Pesnell2012}. It was located near the northwestern limb from SDO's perspective; its tornado-like evolution lasted for about 3 hours during 04:00--07:00 UT on 1 November 2014. AIA takes full-disk images up to $1.5\,R_\odot$ at $0''.6$ spatial scale and 12-s cadence, in seven EUV narrow-band channels spanning a broad range of temperature sensitivities. In this paper, we concentrated on the 171~{\AA} (\ion{Fe}{9}, $\log T = 5.85$) and 304~{\AA} (\ion{He}{2}, $\log T = 4.7$) channels, which displayed most clearly the dynamics of the prominence studied. The ``Ahead'' satellite (hereafter STA) of the Solar Terrestrial Relations Observatory \citep[STEREO;][]{Kaiser2008} captured only the final stage of the tornado, with a single image taken by the Extreme Ultraviolet Imager \citep[EUVI;][]{Howard2008} at each of the four wavelengths, 304, 171, 195 and 284~{\AA}, at about 09:00 UT on 1 November 2014. Since the separation angle between STA and Earth is about 170$^\circ$, the prominence was located near the northeastern limb from STA's perspective.

\subsection{Results}
\begin{figure}
	\centering
	\includegraphics[width=\hsize]{overview.eps}
	\caption{Snapshots of the tornado prominence in the 171 and 304~{\AA} passbands. Co-temporal images from the two passbands are placed side by side and labeled with the same letter but different numerals. On Panel (b2) we superimposed the four virtual slits used to construct time-distance maps, including three horizontal slits (labeled `H1', `H2', and `H3') and a vertical slit (labeled `V'). An animation of the AIA 304 and 171~{\AA} images is available at \url{http://staff.ustc.edu.cn/~rliu/preprint/fig1.mp4}.  \label{fig:overview}}
\end{figure}

\begin{figure}
	\centering
	\includegraphics[width=\hsize]{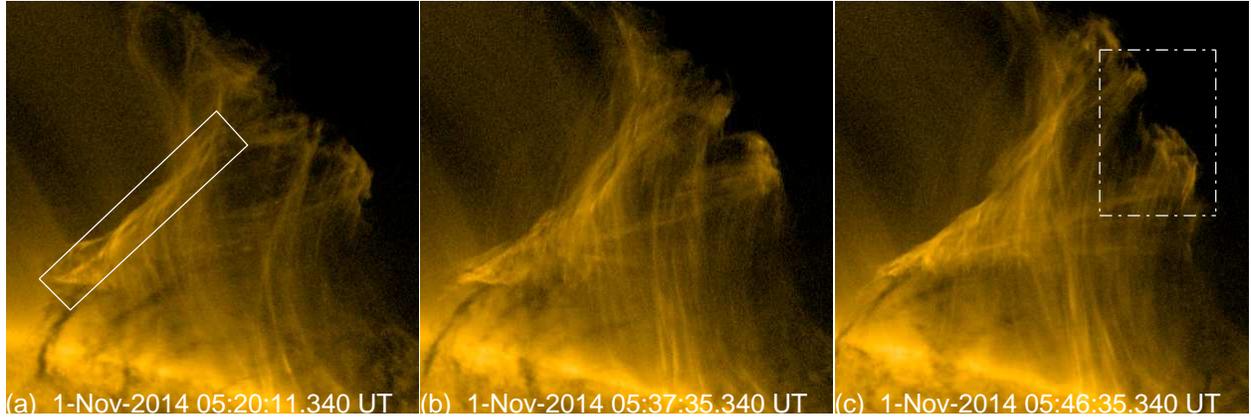}
	\caption{Braided and curtain-like structures in the tornado observed in 171~{\AA}. Images have been rotated by 47 degrees counter-clockwise. The braided structure is marked by a rectangle in Panel (a) and its unraveling process is shown in Panels (b) and (c). The box in (c) indicates the field of view of Figure~\ref{fig:surface}. An animation of AIA 171~{\AA} images is available \url{http://staff.ustc.edu.cn/~rliu/preprint/fig2.mp4}.  \label{fig:braid}}
\end{figure}

\begin{figure}
	\centering
	\includegraphics[width=\hsize]{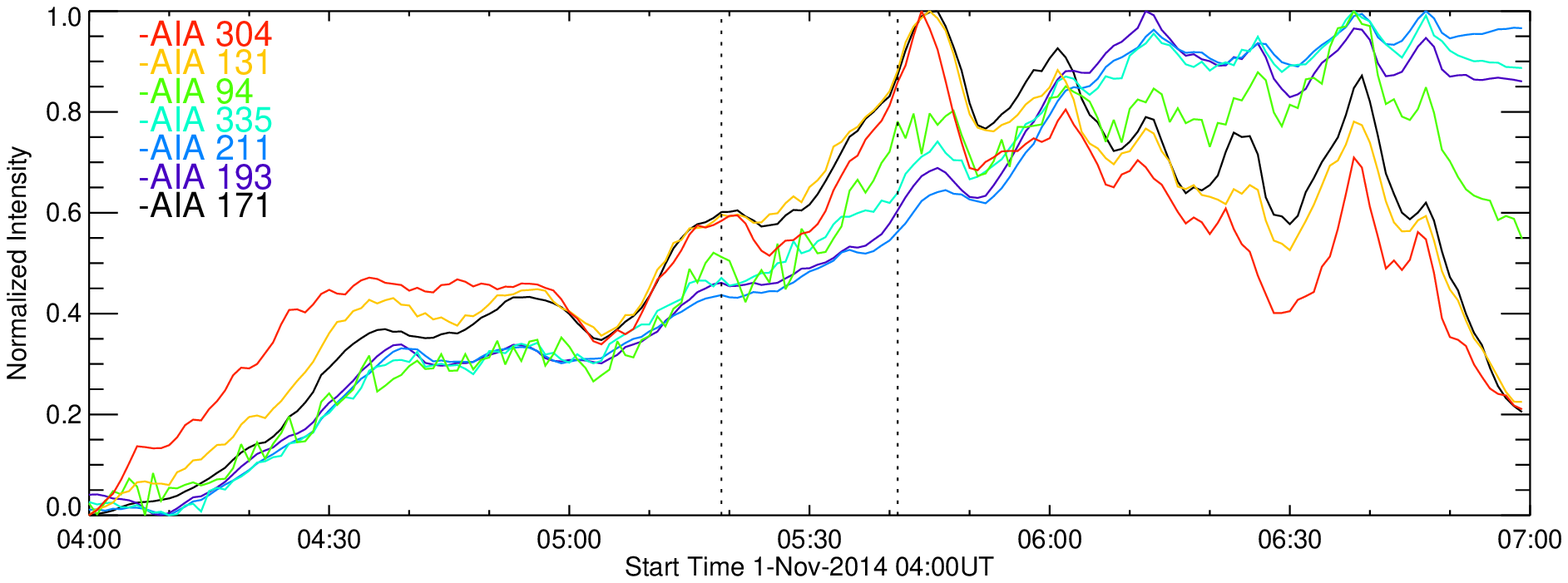}
	\caption{Transient brightening associated with the unraveling of the braided structure. The plot shows temporal variation of the normalized average intensity from the rectangle in Panel (a) of Figure~\ref{fig:braid} with the seven AIA EUV passbands. The vertical dotted lines indicate the time duration during which the braided structure was clearly visible.   \label{fig:braid2}}
\end{figure}

From SDO's perspective, the prominence of interest first appeared on the northeastern limb on 19 October 2014. It was quiescent during the disk passage, until late October 31 when it approached the northwestern limb. Figure~\ref{fig:overview} shows a series of SDO/AIA 171 and 304~{\AA} snapshots showing the key evolution stages of the tornado prominence. Before about 15:00 UT on October 31, the prominence was quiet except for counter-streaming flows \citep{Zirker1998} along its spine. These flows were probably present all the time but better discernible when the prominence was closer to the limb. Starting from about 15:00 UT on October 31 till 02:30 UT on November 1, the prominence body rose upward slowly and formed an arch-shaped structure (Figure~\ref{fig:overview}(b1 and b2), with dark, apparently intertwining threads running along the arch. Within the next two hours, it continued to rise to about 150 Mm above the limb, with the top of the arch slightly kinked due to a slow counterclockwise rotation if viewed from above (Figure~\ref{fig:overview}(c1 and c2)).

\begin{figure}
	\centering
	\includegraphics[width=\hsize]{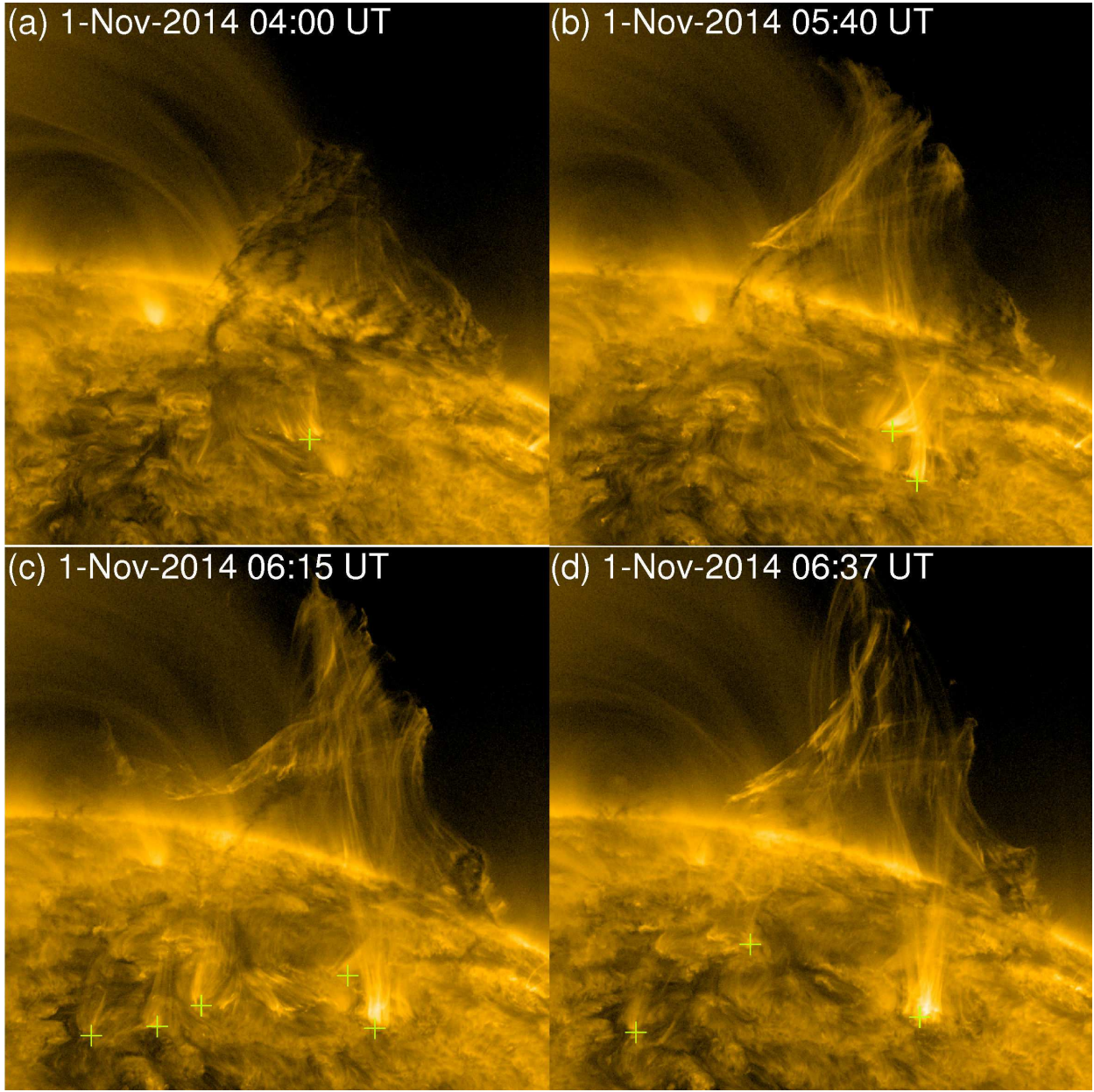}
	\caption{Curtain-like structure formed during the tornado. The crosses mark the footpoints of thread-like loops that constitute the curtain. Images have been rotated by 47 degrees counter-clockwise.  \label{fig:inter}}
\end{figure}

From about 05:00 UT on November 1 onward, the prominence material became brightened in both 171 and 304~{\AA}, indicative of heating. Meanwhile, the arch continued its counterclockwise rotation. A braided structure \citep{Cirtain2013} was visible at the eastern edge of the writhing prominence body (marked by a rectangle in Figure~\ref{fig:braid}(a)). One must keep in mind, however, that this can be formed by the projection of several writhed threads in the optically thin corona. The braided threads appeared to be unraveled later (Figure~\ref{fig:braid}(b--c)), which was associated with a transient brightening in 304, 171, and 131~{\AA} (\ion{Fe}{8}, $\log T = 5.6$) above the `background' enhancement in emission (Figure~\ref{fig:braid2}), and might contribute to the formation of a curtain-like structure (CLS), which was composed of vertically oriented threads (Figure~\ref{fig:braid}(b--c)). Moving westward, the threads appeared to rotate about the vertical in a counterclockwise sense if viewed from above, and at the same time also to rotate slightly in the plane of sky about the top of the prominence (Figure~\ref{fig:overview}(d1, d2); see also the accompanying animation). Material was observed to fall along these loop-like threads, resulting in bright patches at the surface (marked by crosses in Figure~\ref{fig:inter}). These bright patches indicated the footpoints of the threads, which were anchored outside the filament channel, and patches farther away from the channel were seen with the rising of the prominence (see also the animation accompanying Figure~\ref{fig:overview}).

\begin{figure}
	\centering
	\includegraphics[width=\hsize]{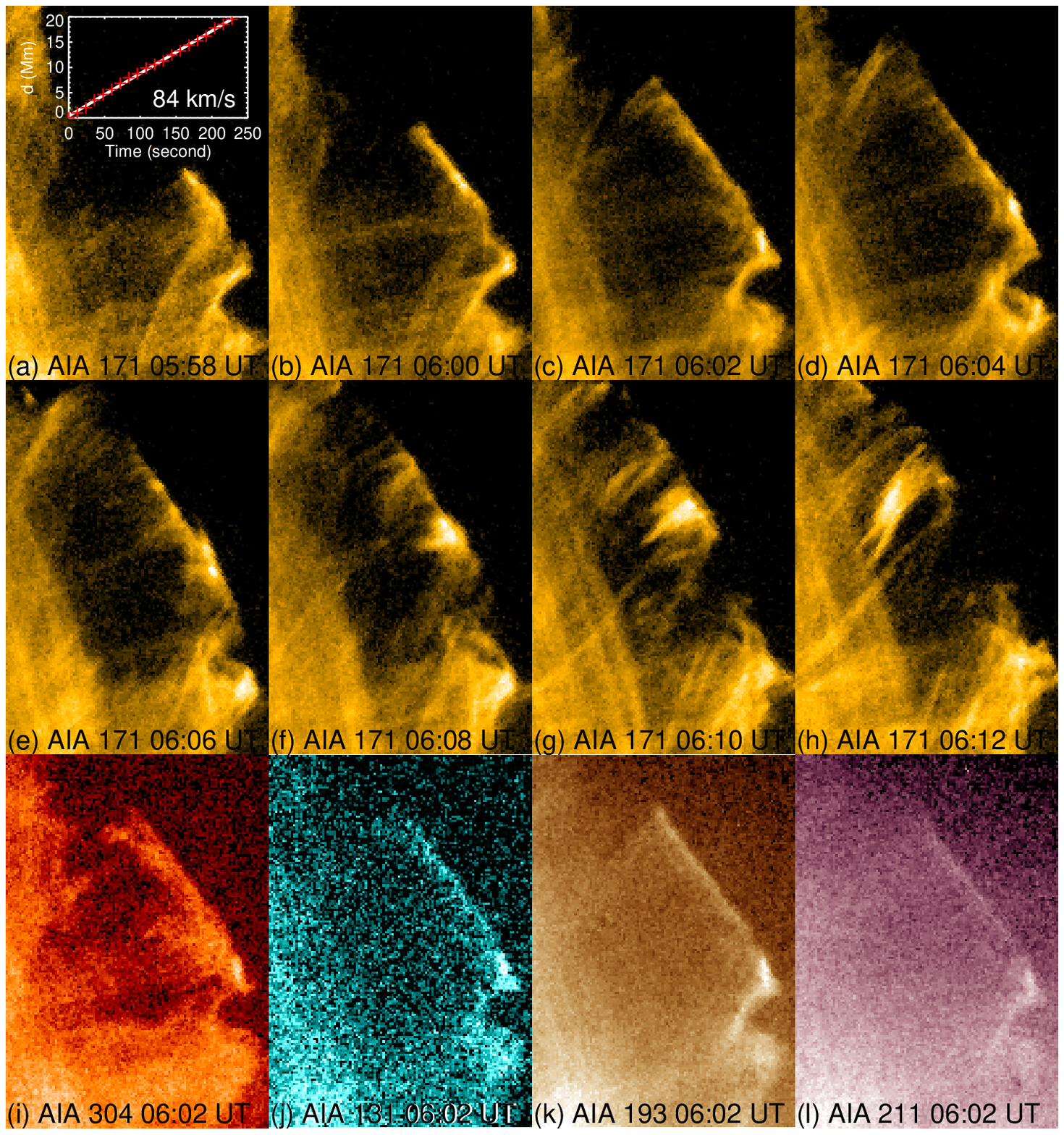}
	\caption{Linear feature at the western edge of the tornado prominence. (a--h) Development and disintegration of the linear feature in AIA 171~{\AA}. (i--l) The linear feature visible in other AIA passbands. The field of view is indicated by the box in Figure~\ref{fig:braid}(c). The inset in Panel (a) plots the time variation of the distance $d$ of the tip of the linear feature from its initial position at 05:58:11 UT. The linear fitting gives a speed of 84 km~s$^{-1}$. \label{fig:surface}}
\end{figure}

Meanwhile, a linear feature at the interface between the prominence and the ambient corona was observed to extend upward at a quasi-constant speed of 84 km~s$^{-1}$ (Figure~\ref{fig:surface}(a--c) and the animation accompanying Figure~\ref{fig:braid}). Like the heated prominence material, the linear feature was most clearly observed in 171~{\AA}, also weakly visible in 304 {\AA}, 131~{\AA} and 193~{\AA} (\ion{Fe}{12}; $\log T=6.2$), but became even weaker in 211~{\AA} (\ion{Fe}{14}; $\log T=6.3$; bottom panels in Figure~\ref{fig:surface}), and nearly invisible in 335~{\AA} (\ion{Fe}{16}; $\log T=6.45$) and 94~{\AA} (\ion{Fe}{18}; $\log T=6.85$). This indicates that the bulk of the filament plasma was heated to about $7\times10^5$ K, the peak response temperature of the 171~{\AA} passband. The extension speed of the linear feature was below the sound speed of $7\times10^5$ K plasma\footnote{$c_s=\sqrt{\gamma k_BT/\mu m_p}$, where $\mu=0.58$ is the mean molecular weight of the fully ionized plasma with 90\% H and 10\% He.}, which is 130 km~s$^{-1}$. The limited resolution prevents us from identifying sub-structures of the linear feature, but later spike-like structures were observed to develop at the prominence-side of the linear feature, which led to the collapse and disappearance of the linear feature within 10 minutes (Figure~\ref{fig:surface}(d--f)). A similar linear feature reappeared at about 06:21 UT and collapsed by 06:30 UT (see the animation accompanying Figure~\ref{fig:braid}).

\begin{figure}
	\centering
	\includegraphics[width=0.9\hsize]{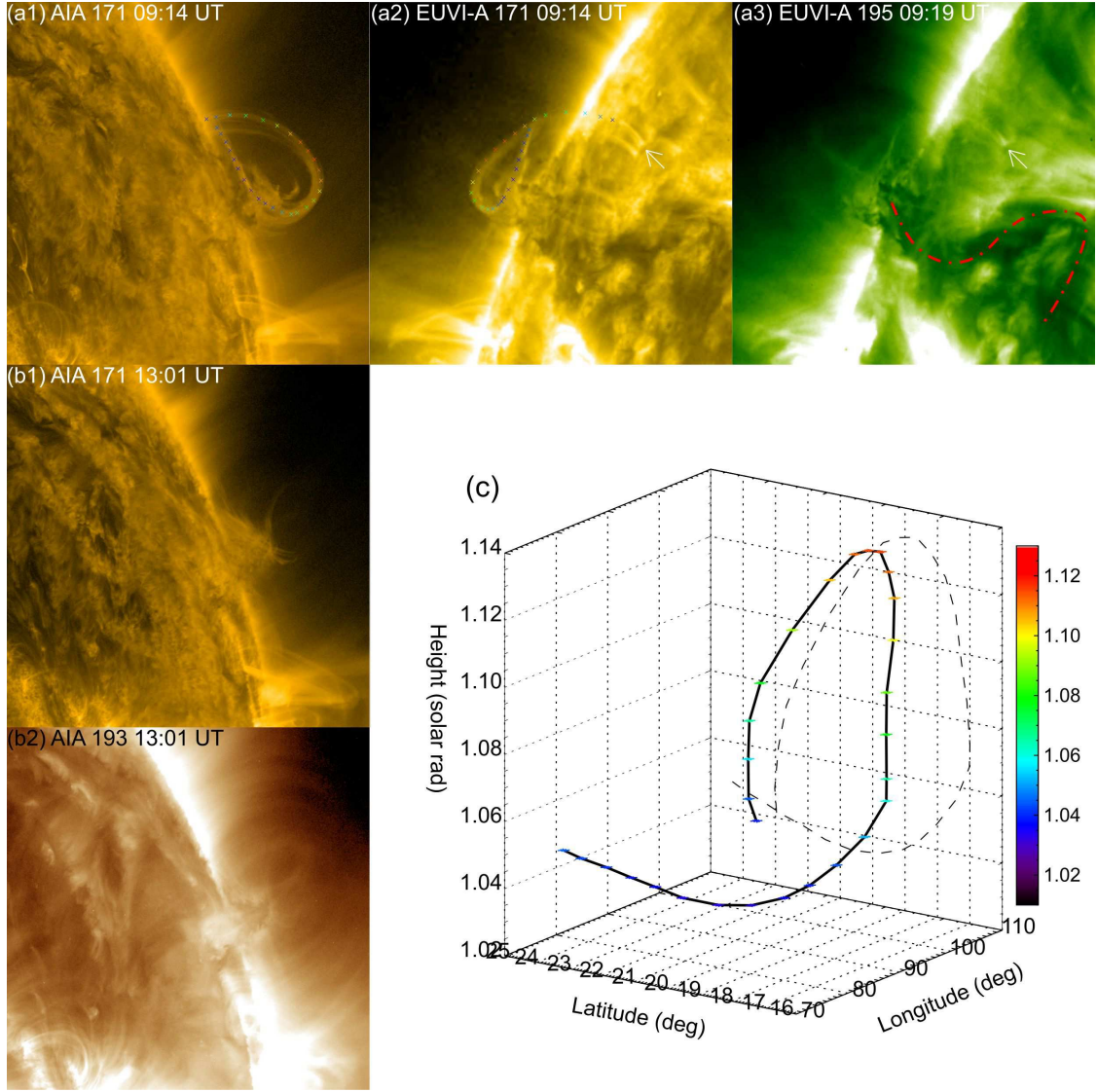}
	\caption{Prominence-cavity system in the wake of the tornado. Top panels (a1, a2, and a3) show the helical structure from the SDO (a1) and STA (a2 and a3) perspectives at approximately the same time. Points chosen for stereoscopic triangulation are marked by crosses in (a1) and (a2), with their reconstructed heights in units of solar radii coded by colors (see the color bar in Panel (c)). The arrows in (a2) and (a3) mark where the top arched section of the helical structure is anchored on the surface. The dash-dotted red curve in (a3) illustrates the filament on the disk. Left panels (a1, b1, and b2) show the prominence from SDO's perspective. (b1) and (b2) were both taken at 13:01 UT on 1 November 2014, four hours after (a1). One can see the pillar-like prominence with a U-shaped ``horn'' on the top in 171~{\AA} (b1), which is located at the bottom a cavity in 193~{\AA} (b2). Panel (c) shows the stereoscopic reconstruction of the helical path of mass flows using the paired SDO/AIA and STA/EUVI images (a1 and a2). Its projection on the local latitude-height plane is shown as a dashed curve to emulate the helical structure above the limb from SDO's perspective.  \label{fig:helical}}
\end{figure}

\begin{figure}
	\centering
	\includegraphics[width=\hsize]{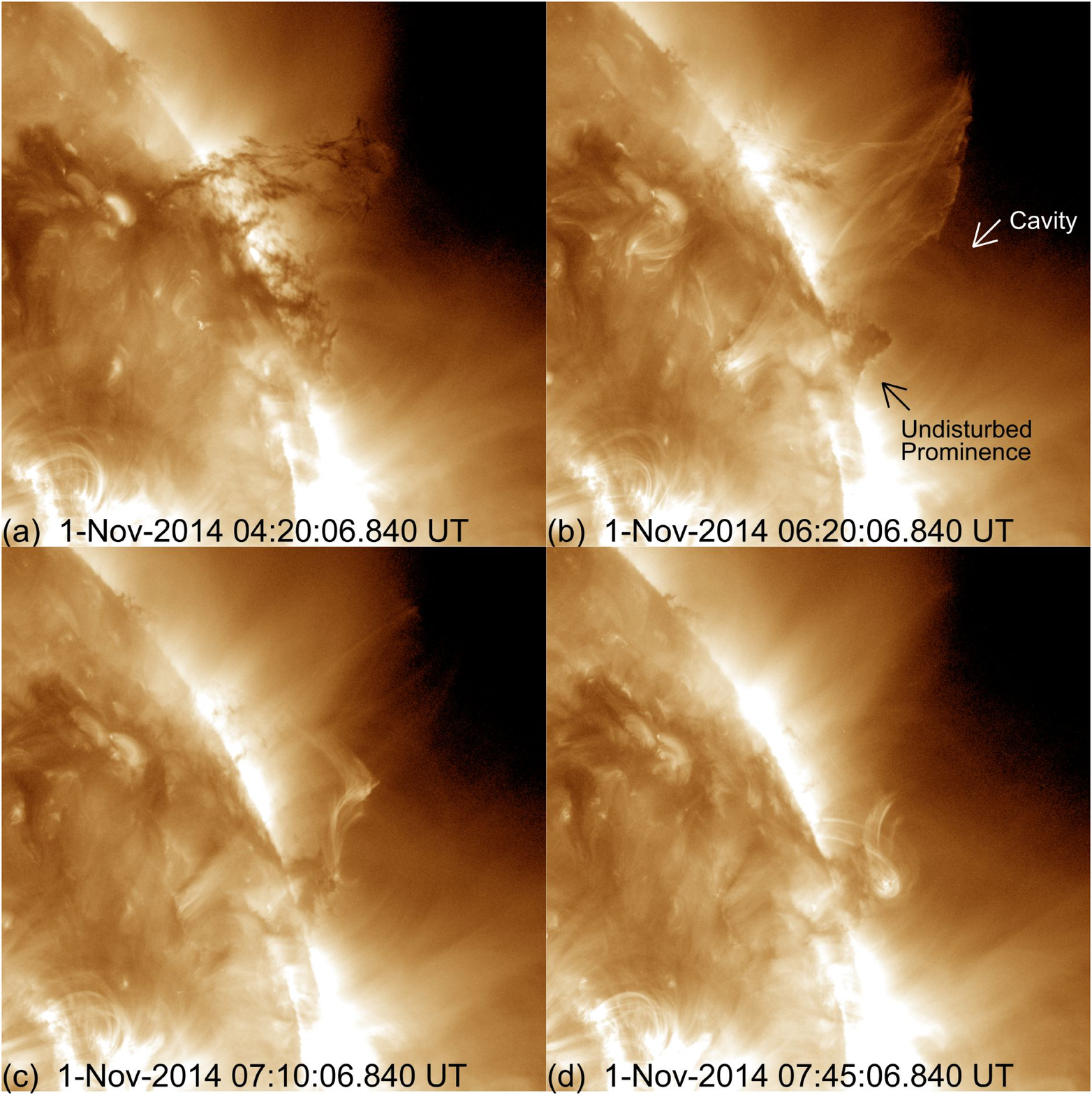}
	\caption{Evolution of the prominence-cavity system (PCS) observed in AIA 193~{\AA}. a) A snapshot of the tornado-like evolution in the eastern section of the prominence; b) undisturbed, western section of the prominence side by side with the tornado; (c--d) helical flows into the PCS by the end of the tornado. \label{fig:cavity}}
\end{figure}

After about 06:00 UT on November 1, The CLS consisting of vertical threads started to fade and gave way to an inner CLS consisting of horizontal threads, with material moving westward along the threads (Figure~\ref{fig:overview}(e1 and e2)). At about 07:30 UT both CLSs disappeared completely. Meanwhile, brightening prominence material flew along a helical trajectory back to the surface (Figure~\ref{fig:helical} and \ref{fig:cavity}(c and d)). Comparing SDO and STA observations, one can see that the top arched part of the helical structure was anchored behind the limb (marked by an arrow in Figure~\ref{fig:helical}(a2 and a3)) and the lower horizontal part was apparently located in the foreground (see also the animation accompanying Figure~\ref{fig:overview}). To confirm this, we used the paired EUVI and AIA 171\,\r{A} images at 09:14 UT (Figure~\ref{fig:helical}(a1 and a2)) to reconstruct the helical trajectory in 3D (Figure~\ref{fig:helical}(c)), with \texttt{scc\_measure} in SolarSoftWare. The reconstruction yields a helical path of about one full turn. Moreover, it is left-handed, in agreement with the sense of the writhing. However, one should keep in mind that this feature was observed with two almost opposite lines of sight. The reconstruction would not be reliable if the feature were surface-like along the line of sight. However, if the helical feature is curve-like, one can estimate the reconstruction error of a single point along the curve due to its finite width $w$ in the plane spanned by the local normals of the two projection surfaces \citep{Inhester2006}. This error could be as large as $\pm8000$ km \citep[][Eq.~(11)]{Inhester2006} if $w$ is set to $2''$, the resolution of EUVI images, but it amounts to only 5\% of the longitudinal extent of the reconstructed helix (28.8 deg or 350000 km on the surface). Thus, we are confident that the helix is left-handed and non-coplanar.  

The helical structure disappeared at about 13:00 UT, but a U-shaped bright ``horn'' remained at the top of the pillar-like prominence in AIA 171~{\AA} (Figure~\ref{fig:helical}(b1)), which is typical of the PCS observed in EUV \citep{Schmit&Gibson2013}. The cavity was better discernible in AIA 193~{\AA} (\ion{Fe}{12}, $\log T = 6.2$; Figure~\ref{fig:helical}(b2) and Figure~\ref{fig:cavity}(b--d)). From the STA perspective (Figure~\ref{fig:helical}(a2 and a3)), this cavity was associated with the western section of the prominence (illustrated by a dash-dotted red curve in Figure~\ref{fig:helical}(a3)), which remained dark in EUV and resided below the helical structure. From SDO's perspective, the cavity became clearly discernible at about 06:00 UT in AIA 193~{\AA} (Figure~\ref{fig:cavity}(b)), presumably when the undisturbed, western section of the prominence was largely aligned along the line of sight, the preferable viewing angle for the observation of cavities. At that time, the cavity was located side by side with the tornado that appeared to have disrupted the eastern section of the prominence.  

Various virtual slits were utilized to construct time-distance maps. Of these slits, four of them are indicated in Figure~\ref{fig:overview}(b2), labeled as `H' or `V', depending on whether their orientation is horizontal or vertical to the prominence. The origin in distance corresponding to the eastern end (bottom) of the horizontal (vertical) slits.

\begin{figure}
	\centering
	\includegraphics[width=\hsize]{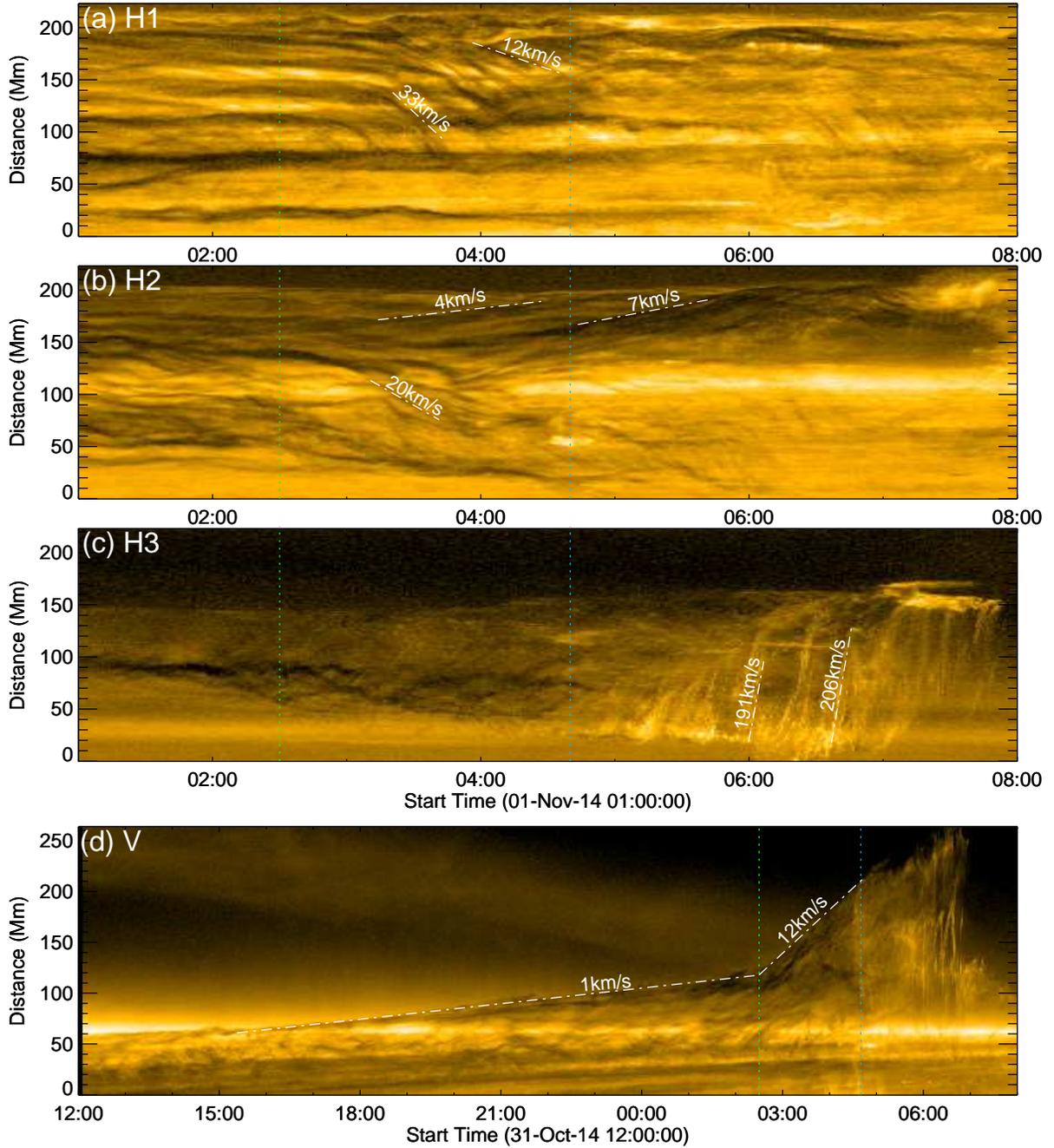}
	\caption{Evolution of the tornado prominence seen through four different slits. The slit locations are marked in Figure~\ref{fig:overview}(b2). Note that the scale of the time axis in (a--c) is different from that in (d). The first vertical dotted line marks the transition from a slow to rapid rise of the prominence at 02:30 UT on 1 November 2014. The second vertical dotted line marks the onset of the tornado and the heating of the prominence at about 04:40 UT. Dash-dotted lines indicate the fitting of various linear features on the time-distance maps.\label{fig:slit}}
\end{figure}

The time-distance map obtained from the vertical slit (Figure~\ref{fig:slit}(d)) shows the slow rise of the prominence at an average speed of $\sim$1 km~s$^{-1}$ before 02:30 UT on 1 November 2014 followed by the sudden transition to a fast rise at about 12 km~s$^{-1}$ during 02:30--04:40 UT. The slow rise speed given by the time-distance map is an overestimate due to the Sun's rotation, which is usually not significant but difficult to gauge \citep{McCauley2015}. The maximum projected height of the tornado reached as large as 250 Mm. Horizontal slits `H1' and `H2' are devoted to the low-lying barbs, while `H3' to the tornado. Prominence material at the H1 height appears to be rather stationary, but during the fast rise period, motions along H1 were dominated by a collective eastward flow at $\sim$10--30 km~s$^{-1}$ (Figure~\ref{fig:slit}(a)), and along H2 by both eastward and westward flows at similar speeds (Figure~\ref{fig:slit}(b)). In the time-distance map obtained from H3 (Figure~\ref{fig:slit}(c)), one can see that the heating of dark prominence material started at about 04:40 UT and that the apparent motions along H3 were dominated by westward flows at about 200 km~s$^{-1}$, due to the counterclockwise rotation about the vertical direction. The opposite flow directions at different heights may correspond to the writhing motion of the prominence body as a whole, but may also likely imply a large-scale circulation of material in this tornado prominence.

\section{Discussion and Conclusion}
The tornado prominence studied in this paper exhibit complex dynamics that is distinct from the tornado studied by \cite{Li2012}. Its evolution can be categorized into three characteristic stages as follows.

\begin{figure}
\centering
\includegraphics[width=\hsize]{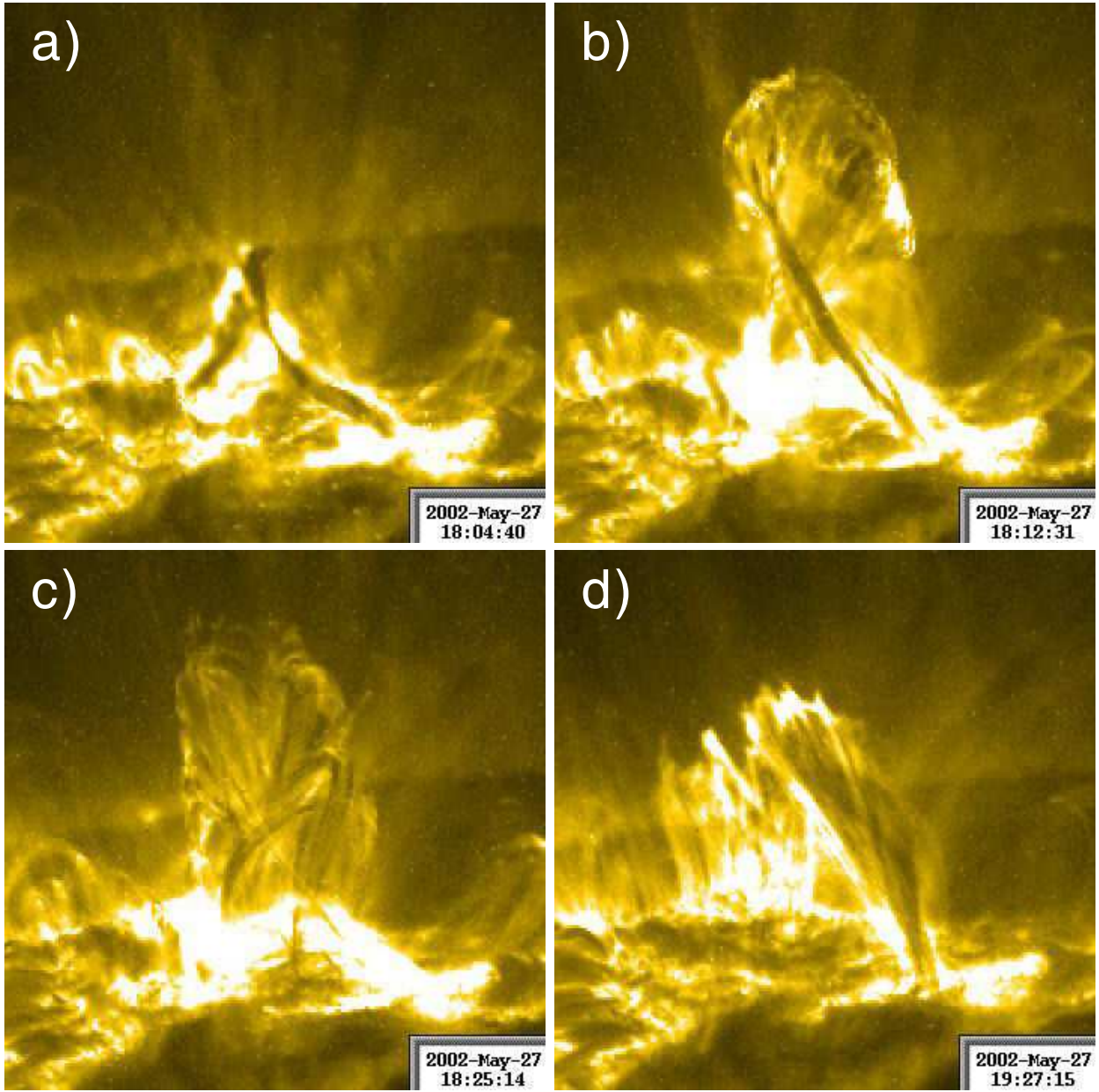}
\caption{Characteristic evolution stages of the confined prominence eruption on 2002 May 27 observed by the Transition Region and Coronal Explorer \citep[TRACE;][]{Handy1999} in 195~{\AA}. Panels (a) and (b) kink stage; (c) disintegration stage; (d) reformation stage. An animation of the TRACE images is available at \url{http://trace.lmsal.com/POD/movies/T195_020527_18M2.mov}.)\label{fig:02may27}}
\end{figure}

\begin{enumerate}

\item \textbf{Kink stage.} Before the tornado, the prominence develops an arch-shaped structure through a slow rise at $\sim\,$1~km~s$^{-1}$, which is frequently observed during the initial stage of prominence eruptions \citep[e.g.,][]{Liu2012}. The onset of the tornado is marked by a sudden transition to a rapid rise over 10~km~s$^{-1}$, which is associated with a left-handed writhing of the arch (Figure~\ref{fig:overview}). This suggests that the prominence is kink unstable. 

\item \textbf{Disintegration stage.} The prominence resembles terrestrial cyclones, featured by a maelstrom of rotational and draining motions of prominence material. The heating of the prominence material to a few $10^5$ K during this stage might be contributed by the conversion of magnetic twist to writhe, the unraveling of the braided structure in the prominence, or the interactions/reconnections between the prominence field and the overlying coronal field. We highlight below interesting structures identified in the tornado.

\begin{itemize}
	\item \textbf{Curtain-like structure.} Two bright CLSs are observed during the tornado, both composed of loop-like threads (Figure~\ref{fig:overview}). Here we focus on the CLS consisted of vertically oriented threads, which appear to rotate about the vertical in a counterclockwise sense if viewed from above, consistent with the left-handed writhing of the arch-shaped structure. Heated filament material is observed to drain down along these vertical threads, resulting in brightening patches at the surface, located apparently outside the filament channel; patches farther away from the channel are seen with the rise of the prominence (Figure~\ref{fig:inter}). To understand the implication of this observation, we resort to the generally accepted physical picture of filaments, in which the dense filament material suspended in the corona is supported against gravity by a highly non-potential core field, either twisted or sheared, in the filament channel. This core field is enclosed by an overlying arcade of much more potential field. During the quiet period, the overlying field contains no prominence material, hence is not easily discernible; during the tornado, however, heated prominence material flows along curved paths reminiscent of overlying field lines rather than ballistic trajectories, falling towards the surface far away from the filament channel. This can only happen if reconnection occurs, joining these two topologically distinct regions, so that the prominence material is allowed to be transported from the core to the overlying field. Thus, the threads constituting the CLS represent new magnetic connections. The renowned kink-unstable prominence on 2002 May 27 \citep{Ji2003,Alexander2006,Torok&Kliem2005} also develops a similar CLS after the saturation of the helical kink (Figure~\ref{fig:02may27}(c)). This is an important signature of reconnection between the flux rope flux and the overlying flux, as carefully modeled by \citet{Hassanin&Kliem2016}. 
	
	\item \textbf{Braided structure.} A braided structure composed of the intertwining prominence threads appears during the initial phase of the tornado (Figure~\ref{fig:braid}). Its subsequent unraveling might contribute to the formation of the CLS of vertical threads and be associated with heating, as demonstrated by a transient brightening around the region of the braided structure in EUV (Figure~\ref{fig:braid2}).
	
	\item \textbf{Interface anomaly.} At the edge of the tornado prominence a linear feature is observed repeatedly to extend upward at a constant sub-acoustic speed (Figure~\ref{fig:surface}). It soon collapses with the development of spike-like structures, which might suggest that a certain surface instability is in progress. On the other hand, the linear feature could be a collection of the outermost points of the threads that form the CLS. Its disintegration hence results from the different dynamical evolution of the individual threads. In that case, however, the endpoints of these threads would have to appear successively at a uniform speed to reproduce the observation, which we deem as a less likely scenario.
\end{itemize}

\item \textbf{Reformation stage.} A helical structure, outlined by mass flow, appears in the wake of the tornado (Figure~\ref{fig:helical}). It is relatively stable lasting about six hours before its disappearance in both 171 and 304~{\AA} (see the animation accompanying Figure~\ref{fig:overview}), presumably due to mass drainage. In projection, the bulk of the remaining prominence material resides at the bottom of the helical structure, which is consistent with the classical flux-rope model of quiescent prominences \citep{Low&Hundausen1995}. Similarly in the 2002 May 27 event, a much less twisted prominence survived the eruption (Figure~\ref{fig:02may27}(d)). Numerical simulations by \citet{Hassanin&Kliem2016} demonstrate the reformation of the flux rope via reconnection in the vertical current sheet between the two original flux rope legs. In our case, however, there was no clear evidence of reconnection during this stage. Considering that the observed PCS is located to the immediate west of the tornado (Figure~\ref{fig:overview} and \ref{fig:cavity}), we suggest that only the eastern section of the original prominence rises into the arch-shaped structure and experiences the tornado-like evolution, while the western section, though probably disturbed yet not disrupted by the tornado, settles down and reorganizes into a coherent PCS.

\end{enumerate}

\begin{figure}
	\centering
	\includegraphics[width=\hsize]{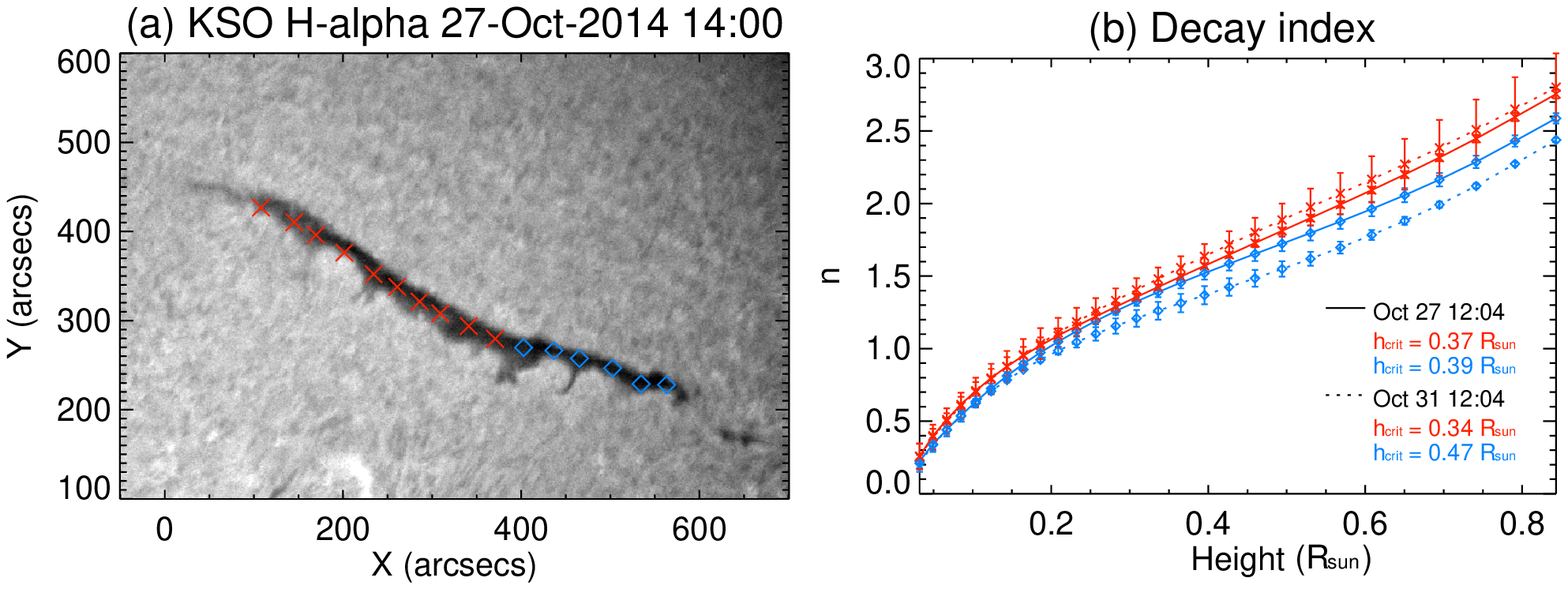}
	\caption{Decay index of the background field. Left: The filament of interest as observed in H$\alpha$ by the Kanzelh\"{o}he Solar Observatory (KSO) on 2014 October 27. The red crosses (blue diamonds) indicate manually picked points along the eastern (western) section of the filament spine, above which the decay index $n$ at different heights is calculated. These points are updated with time, taking into account the differential rotation of the Sun. Right: The mean decay index profile for the eastern (western) section is shown in red (blue), with error bars indicating the standard deviation, at two different times as indicated in the legend. The height $h_\mathrm{crit}$ corresponding to the theoretical threshold of the torus instability ($n_\mathrm{crit}=1.5$) for each individual profile is also given in the legend.  \label{fig:decay}}
\end{figure}

To understand why the prominence's two sections behave differently, we calculated the decay index $n=-d\log B_h/d\log h$ \citep{Kliem&Torok2006} of the background field, using the \texttt{pfss} package in SolarSoftWare,  which takes into account the evolving field on the full sphere by assimilating magnetograms into a flux-dispersal model \citep{Schrijver&DeRosa2003}, and yields the coronal field with a potential-field source-surface model. In our calculation $B_h$ is the `horizontal' field component perpendicular to the radial component $B_r$. It is found that the profile of $n$ with height differs significantly between the eastern and western section of the prominence as it evolves toward the time of the tornado, and that the western section is better confined by the background field, with the blue profile being consistently below the red one (Figure~\ref{fig:decay}). Hence, it is not surprising that the two sections evolve independently to a certain degree. For the eastern section, the theoretical threshold of the torus instability, $n_\mathrm{crit}=1.5$, corresponds to a height as high as 0.34 $R_\odot$ (236 Mm) above the surface on October 31, hours prior to the tornado, while the prominence only reached about 150 Mm above the limb, or a projected height of about 210 Mm (Figure~\ref{fig:slit}(d)), when it still maintained a coherent shape before being disintegrated by the tornado. That the prominence remained torus stable may account for its failure to erupt as a coronal mass ejection. 

One should keep in mind that the ``true'' critical decay index might deviate from 1.5, which is derived for an idealized current ring \citep{Kliem&Torok2006}. In MHD simulations $n_\mathrm{crit}$ is found to be in the range [1.4--1.9] \citep[][and references therein]{Zuccarello2016}. Hence, $n=1.5$ can serve as a reasonable reference number in the above analysis. \citet{Zuccarello2016} noticed that a discrepancy between theoretical predictions and observational results may result from whether the decay index is evaluated at the top of the prominence, where $n\approx1.1$ in their simulations, or at the apex of the flux rope axis, where $n\approx1.4$. However, when a prominence is highly dynamic, as in the present case, prominence material often moves around along field lines, rather than reside quietly at the dipped portions of field lines, which is expected to significantly reduce the aforementioned discrepancy.   

To conclude, the tornado of interest is governed globally by the helical kink instability, and affected locally, by surface instabilities. The prominence is disintegrated via interactions between the prominence field and the overlying coronal field, which resulted in a curtain-like structure consisting of thread-like loops anchored outside the filament channel. The remaining heated prominence material flows along a left-handed helical path into the PCS associated with the western section of the prominence that survives the tornado. The western section hence bears the same sense of helicity as the eastern section which displays left-handed writhing. The cyclonic behavior in the present event is dominated by rotational motions about the vertical, whereas the tornado studied by \citet{Li2012} features swirling motions in the plane of sky. The distinction may signal different physical processes involved but could also result from different viewing angles, which will be a topic for future investigation. In any case, we propose that the kink instability and the magnetic reconnection between the prominence field and the overlying field may play a significant role in some of the solar tornadoes, in which numerous filament threads undergoing untwisting/writhing motions may collectively create a cyclonic illusion. 

As a final remark, we point out a possibility that the writhing motion could be caused by the misalignment between the sheared ambient field and the current carried by the flux rope, due to the rise of the rope \citep{Isenberg&Forbes2007}, although this effect is expected to be insignificant in our case because of weak magnetic shear in the quiet Sun. Moreover, the CLS might not be unique to confined eruptions of kink-unstable prominences. It might also be present in successful prominence eruptions without writhing, such as the well-studied 7 June 2011 event, during which the impact sites of falling filament material were far away from the source region \citep{Reale2013,Gilbert2013,Carlyle2014}. Although many descending filament blobs in that event were apparently not fully guided by field lines but significantly affected by inertia and gravity, \citet{vanDriel-Gesztelyi2014} found an unambiguous episode of reconnection between the eruptive magnetic structure and a neighboring AR, which manifests as the redirection of the filament plasma toward remote magnetic footpoints and the local heating of plasma at the reconnection region to a temperature mainly covered by the 171~{\AA} channel. This process shares similarity with the formation of the CLS in the present event and that modeled by \citet{Hassanin&Kliem2016}.

\acknowledgments We thank B.~Kliem for helpful comments, and the anonymous referee for constructive suggestions. RL acknowledges the support from the Thousand Young Talents Program of China and NSFC 41474151. YW acknowledges the support from NSFC 41131065 and 41574165. This work was also supported by NSFC 41421063, CAS Key Research Program KZZD-EW-01-4, and the fundamental research funds for the central universities.

\end{document}